\documentclass[aps,twocolumn,showpacs,superscriptaddress]{revtex4}
\usepackage{graphicx}
\usepackage{dcolumn}
\usepackage{bm}
\usepackage{amssymb}
\begin{document}
\preprint{APS}
\title{Anomalous dynamical scaling in anharmonic chains \\ and plasma models with 
multiparticle collisions}
\author{Pierfrancesco Di Cintio}
\email{pdicintio@unifi.it}
\affiliation{Dipartimento di Fisica e Astronomia and CSDC, Universit\'a di Firenze, 
via G. Sansone 1, I-50019 Sesto Fiorentino, Italy}
\affiliation{Istituto Nazionale di Fisica Nucleare, Sezione di Firenze, via G. Sansone 1, I-50019 Sesto Fiorentino, Italy}
\author{Roberto Livi}
\affiliation{Dipartimento di Fisica e Astronomia and CSDC, Universit\'a di Firenze, 
via G. Sansone 1, I-50019 Sesto Fiorentino, Italy}
\affiliation{Istituto Nazionale di Fisica Nucleare, Sezione di Firenze, via G. Sansone 1, I-50019 Sesto Fiorentino, Italy}
\author{Hugo Bufferand}
\affiliation{Aix-Marseille Universit\'e, CNRS, PIIM, UMR 7345, F-13397 Marseille Cedex 20, France}
\author{Guido Ciraolo}
\affiliation{CEA, IRFM, F-13108 Saint-Paul-lez-Durance, France}
\author{Stefano Lepri}
\affiliation{Consiglio Nazionale delle Ricerche, Istituto dei Sistemi Complessi via Madonna del piano 10, I-50019 Sesto Fiorentino, Italy}
\affiliation{Istituto Nazionale di Fisica Nucleare, Sezione di Firenze, via G. Sansone 1, I-50019 Sesto Fiorentino, Italy} 
\author{Mika J. Straka}
\affiliation{IMT Institute for Advanced Studies Lucca, Piazza S. Francesco 19, I-55100 Lucca, Italy}
 \date{\today}
\begin{abstract}
We study the anomalous dynamical scaling of equilibrium correlations in one dimensional systems.
Two different models are compared: the Fermi-Pasta-Ulam chain with cubic and quartic
nonlinearity and a gas of point particles interacting stochastically 
through the multiparticle collision dynamics. For both models -that admit three conservation laws- by means of detailed numerical simulations we verify the predictions of nonlinear fluctuating hydrodynamics for the structure factors of density and energy fluctuations
at equilibrium. Despite this, violations of the expected scaling in the currents
correlation are found in some regimes, hindering the observation of the asymptotic
scaling predicted by the theory. In the case of the gas model this crossover
is clearly demonstrated upon changing the coupling constant.
\end{abstract}
\maketitle
\section{Introduction}
Low-dimensional [i.e., one- (1D) and two-dimensional (2D)] systems where total size, energy 
and momentum are the only conserved quantities, typically exhibit anomalous relaxation
and transport properties \cite{2003PhR...377....1L,1986FoPh...16...51C,2007EPJST.151...85B,2008AdPhy..57..457D}. Under these conditions the standard hydrodynamic 
description fails, because the transport coefficients are ill-defined in 
the thermodynamic limit.  For instance, in 2D systems the 
heat conductivity $\kappa$ exhibits a logarithmic divergence with the system size $N$, while in 1D systems
it turns into a power-law 
\begin{equation}
\kappa \sim N^{\gamma}.
\end{equation}
The value of $\gamma$ has been estimated by numerical simulations and various theoretical approaches for different systems. 
For a generic non integrable and nonlinear 1D system it is assumed that the scaling exponent takes the
universal value $\gamma = 1/3$, indicating non-diffusive (anomalous) heat conduction.
Numerical evidence has been found for several systems like the hard 
point gas (HPG), the hard point chains (HPC) (both with alternate masses),
and the Fermi-Pasta-Ulam (FPU) $\alpha + \beta$
model (a Hamiltonian chain of nonlinearly coupled oscillators, interacting by a leading cubic nonlinearity \cite{2003PhR...377....1L}). 
The universality of the $\gamma$ exponent has been predicted by different 
theoretical arguments (e.g., see Refs. \cite{2002PhRvL..89t0601N,2006PhRvE..73f0201D}).
More recently, a complete description has been put forward within  
the Nonlinear
Fluctuating Hydrodynamics (NFH) approach, proposed independently by van Beijeren \cite{2012PhRvL.108r0601V} and 
Spohn \cite{2014JSP...154.1191S,2015arXiv150505987S}. These authors have shown that the statistical properties of 1D nonlinear
hydrodynamics with three conservation laws (e.g. total energy, momentum and number of particles) are essentially described by the fluctuating Burgers equation for the field $\Psi(x,t)$ with white noise $\mathcal{Z}$
\begin{equation}\label{burgers}
\partial_t\Psi+c\Psi\partial_x\Psi=\nu\partial_{xx}^2\Psi+\sqrt{2\nu}\partial_x\mathcal{Z},
\end{equation}
where $\nu$ is a diffusion coefficient and $c$ the scale velocity of advection. Equation (\ref{burgers}), on its side, can be mapped onto the well-known Kardar-Parisi-Zhang
(KPZ) equation for the stochastic growth of interfaces \cite{1986PhRvL..56..889K}. By exploiting this formal equivalence between different problems,
one can conclude that anomalous relaxation and transport in 1D systems can be
traced back to the superdiffusive behavior of density fluctuations imposed by the
constrained dynamics.\\
\indent However, there exist counterexamples of models with 
the three aforementioned conservation laws that depart
from such universal behaviour for different intrinsic reasons. For instance,
integrable models, like a chain of harmonic oscillators and the Toda lattice,
exhibit ballistic transport (i.e., $\kappa \sim N$), since energy is transmitted through
the chain by the undamped propagation of eigenmodes (phonons and solitons, 
respectively \cite{2012PhRvE..85d1118K,1967JPSJ...23..501T}). Moreover, for systems in which the local symmetry of particle displacements with respect to the equilibrium
position is restored, the exponent $\gamma$ takes higher values, e.g. $2/5$ or $1/2$,  
depending on the model at hand. It has been argued that this case belongs to a
different 
universality class. This has been tested numerically in the FPU-$\beta$ model, where the displacement symmetry is a straightforward consequence
of a purely quartic nonlinearity \cite{Lepri03,2005PhRvE..72c1202L,Delfini08c}, 
in the hard-point-chain model at zero pressure \cite{2011JSMTE..03..028P}
and in the FPU-$\alpha + \beta$ model, where the displacement symmetry
can be imposed by applying a suitable pressure at the chain boundaries \cite{2008JSP...132....1L}.
It has been further pointed out that such a case
corresponds to the special thermodynamic condition, for which the specific heat
capacities at constant volume and pressure are equal \cite{2015PhRvE..91c2102L}.
A further remarkable example belonging to this class is provided by the harmonic chain
subject to a conservative noise, where deterministic dynamics coexist with
random collisions among oscillators preserving momentum and energy. In this
case, it has been rigorously proved that $\gamma = 1/2$ \cite{BBO06,2008JSP...133..417B,Lepri2009}.\\
\indent An interesting example of normal conduction in nonlinear systems is the rotor chain
that can be thought as the 1D dynamical version of the $XY$-model. Numerical studies
\cite{2000PhRvL..84.2144G,2014PhRvE..90a2124D} and recent calculations based on the NFH approach \cite{Das2014b,2015arXiv150505987S,2015JSMTE..08..028M} predict finite
$\kappa$ for such a model in the thermodynamic limit. This is
not surprising for a twofold reason: {\sl (i)} in the rotor chain, only energy and momentum are
conserved quantities, while the "length" is not, because the cosine-like
interaction  potential depends on angle-variables, whose relative phase can change
by $\pm 2\pi$ without any consequence on the dynamics, and {\sl (ii)} due to the boundedness
of the potential, at any finite value of the energy density there are energy fluctuations that 
allow some rotors to overtake the potential  barrier and produce spontaneously
localized excitations (the so-called {\sl rotobreathers}), that behave like effective scatterers 
for long-wavelength (i.e. hydrodynamic) fluctuations, thus restoring conditions of normal diffusion.\\
\indent It also deserves to be mentioned that both classical and quantum chains with long-range couplings always show anomalous behaviour in energy transport. In this case, the deviation from the normal 
heat diffusion as described by the classical Fourier law has to be attributed to the non-local nature of the interaction \cite{2014Natur.511..198R}, rather than to the dimensionality of the system. In particular, such models appear to trap large fractions of the total energy into one or more degrees of freedom for long times \cite{2005NYASA1045...68P,2015PhRvE..91c2927M}, thus yielding  subdiffusive transport.\\ 
\indent All of these results indicate that deviations from universality
may emerge in 1D transport and range from normal diffusion to ballistic transport 
depending on specific additional symmetries or dynamical mechanisms at work.\\
\indent Recently, this complex scenario turned to a true puzzle, because careful numerical experiments
reported that normal heat conductivity can be observed in models, in which the universal
exponent $\gamma = 1/3$ is expected to hold \cite{2013PhRvE..87c2153C,2014PhRvE..90c2134C}. The main claim of the authors of these
works is that, at variance with the predictions of NFH and of previous
theoretical approaches, normal heat conductivity should characterize all 1D lattice
models with asymmetric potentials. Such a situation has been observed
for the FPU-$\alpha + \beta$ model and for chains of oscillators coupled through a
Lennard-Jones potential \cite{2005EPJB...47..549L}. The authors provided an explanation of their 
results by arguing that the scattering of long-wavelength phonons could be 
attributed to the spontaneous formation of macroscopic mass gradients
along the chain, typical of asymmetric interaction potentials acting between 
nonlinear oscillators \cite{2012PhRvE..85f0102Z}. The same authors also speculated that the exponent $\gamma=1/3$
is observed  in the HPG,  because in a gas-like model the
asymmetry of the interaction is much less effective in producing macroscopic density fluctuations.
A similar conjecture was invoked to explain why the exponent $\gamma=1/3$ is
observed numerically in models with asymmetric potential when
some parameter, e.g. the energy or the strength of the nonlinearity, is varied. For instance, in the FPU-$\alpha + \beta$ model 
the universal anomalous scaling observed for large values of the energy seems to turn to normal
transport conditions in the low energy regime.
These puzzling results were confirmed also by numerical simulations performed 
in non-equilibrium conditions, by studying stationary
states in the presence of heat baths acting at the chain boundaries \cite{2014PhRvL.112m4101I}.\\ 
\indent In view of these findings, further numerical studies were triggered.  Das, Dhar and
Narayan \cite{2013JSP...tmp..242D}; Wang, Hu, and Li \cite{2013PhRvE..88e2112W}, and Savin and Kosevich \cite{2014PhRvE..89c2102S} repeated the same
kind of simulations and concluded that finite heat conductivity is actually a
finite size effects and does not correspond to the expected universal scaling behavior 
that should be attained in the thermodynamic limit. In practice, the seemingly normal 
diffusion observed in certain dynamical regimes of nonlinear lattices
with asymmetric interaction should crossover to the power law scaling $\gamma = 1/3$ for sufficiently large sizes 
and sufficiently long times.\\
\indent As a matter of fact, it is still quite difficult to check
the crossover in this class of lattice models, because finite size effects may 
change sharply by orders of magnitude when a model parameter, like energy or strength
of nonlinear coupling, is varied. As we are going to discuss, this is a major drawback of nonlinear oscillator
models. In this paper, we explore this matter further by studying the FPU-$\alpha + \beta$ model and showing that predictions
of the NFH are verified for the structure factors associated to density fluctuations in a range
of parameter (i.e. energy) values, where the thermal conductivity $\kappa$ exhibits the
cross-over from anomalous (low-energy) to normal (high-energy) transport.\\
\indent From the side of systems of particles interacting via long-range forces (e.g. plasmas or self-gravitating systems), the problem of the cross-over from normal to anomalous diffusion is even more complicated, as such systems live for long times in states that are out of equilibrium (the so-called quasi stationary states \cite{2006PhyA..365..102C}). In addition, the dynamics of plasmas (as well as of gravitational systems) is principally dominated by {\it mean field} effects rather than by inter-particle collisions due to the long-range nature of the $1/r^2$ force. Moreover, the large number of particles in such systems forces to naturally adopt a description in the continuum collisionless limit in terms 
of the phase-space distribution function $f(\mathbf{r},\mathbf{v},t)$ through the collisionless Boltzmann or Vlasov equation (CBE) \cite{1982A&A...114..211H} 
\begin{equation}\label{vlasov}
\partial_t f+\mathbf{v}\cdot\nabla_{\mathbf{r}}f+\mathbf{F}\cdot\nabla_{\mathbf{v}}f=0
\end{equation}
with self-consistent fields $\mathbf{F}$. Therefore, the numerical modeling of long-range interacting systems is usually carried out with schemes based on the CBE, such as the widely used particle-in-cell (PIC) \cite{2008CNSNS..13..204B}. However, the contribution of collisions to the energy transport is not negligible in certain environments (e.g. hot cores in tokamak fusion plasmas \cite{2000SPP.....7.....S} and dense galactic nuclei \cite{2015ApJ...804...52M}).
\indent In this work we also study a simplified model of a 1D collisional plasma of particles interacting via an effective Coulomb-like interaction based on the multiparticle collision (MPC) scheme, inspired by phenomenological models adopted in plasma physics \cite{2010JPhCS.260a2005B,2013PhRvE..87b3102B}. MPC simulations, previously used to study hydrodynamic correlations \cite{2002PhRvE..66c6702A,2012PhRvE..86e6711H,2015arXiv150807157V} and the fluctuation-dissipation theorem \cite{2011PhRvL.106u0601B} in fluid dynamics, have the advantage of being carried out in a gas-like model, where the typical excitations yielding large density fluctuations typical of lattice models can be neglected.\\
\indent The rest of the paper is structured as follows. In Section. \ref{fpu} we introduce the most important quantities that will be
used to study the results of the numerical simulations in the context of the Fermi-Pasta-Ulam system, while in Section. \ref{mpc}, we describe the MPC plasma model and we show that it allows a better control of the crossover region, confirming that normal heat conductivity
is due to a finite size effect. Moreover, we discuss the possible implementation of the MPC method in standard mesh-based plasma codes and we study the interplay between collisions and mean-field effects. Finally, in Section. \ref{discussion} we summarize and discuss a possible interpretation
of the similarities and analogies of the models analyzed in this paper.
\section{FPU-$\alpha+\beta$ chain}\label{fpu}
\subsection{Model}
We consider a system of $N$ particles of equal mass $m = 1$, arranged in a one dimensional lattice. Let $x_i$ be the position of the $i$-th particle. Assuming that interactions are restricted to nearest-neighbor pairs, the equations of motion for such a system read
\begin{equation}\label{model}
\ddot{x}_i=-F_i+F_{i-1};\quad F_i=-V^\prime(x_{i+1}-x_i)
\end{equation}
where $V^\prime(z)$ is a shorthand notation for the first derivative of the the inter-particle potential $V$ with respect
to $z$. The particles are confined in a simulation box of length $L$ with periodic boundary conditions
\begin{equation}
x_{i+N}=x_{i}+L.
\end{equation}
The FPU-$\alpha + \beta$ potential, in suitable units, is
\begin{equation}
\label{pot}
V(z)=\frac{1}{2}(z-a)^2+\frac{\alpha}{2}(z-a)^3+\frac{\beta}{2}(z-a)^4.
\end{equation}
Throughout this paper we take units such that $\beta=1$ and we only consider cases in which $0<\alpha\ll 1$, so the resulting potential is always single welled. Note that FPU-$\alpha$ models (i.e., $\beta=0$) are excluded from this discussion as they admit run-away solutions for long integration times.\\
\indent Considering a periodic simulation interval with $L=Na$, where $a$ is the equilibrium distance between two neighboring oscillators, we can set $x_i = ia + u_i$, $u_i$ being the displacement from the equilibrium position. With such a choice, widely used in the literature \cite{2007JSMTE..02....7D}, Eq. (\ref{model}) becomes
\begin{eqnarray}\label{edm}
\ddot{u}_i&=&u_{i+1}+u_{i-1}-2u_i+\nonumber\\
&+&\alpha\left[(u_{i+1}-u_i)^2-(u_{i-1}-u_i)^2\right]+\nonumber\\
&+&(u_{i+1}-u_i)^3-(u_i-u_{i-1})^3,
\end{eqnarray}
and the periodic boundary condition now reads $u_{N+i}=u_i$. At site $i$, the local particle energy is defined as 
\begin{equation}
\mathcal{E}_i=\frac{1}{2}\dot{u}^2_i+\frac{1}{2}(u_{i+1}-u_i)^2+\frac{\alpha}{3}(u_{i+1}-u_i)^3+\frac{1}{4}(u_{i+1}-u_i)^4.
\label{locen}
\end{equation}
Note that the inter-particle distance $a$ disappears from
the equations above thus being a completely arbitrary quantity.\\
\indent In order to study the dynamical scaling of the correlations of the energy, and relate them to the thermal conductivity, we introduce the Green-Kubo formula \cite{kersonhuang}
\begin{equation}\label{greenkubo}
\kappa=\frac{D}{k_BT^2N}\int_0^\infty c_J(t){\rm d}t,
\end{equation} 
written here for a finite system of size $N$. In the equation above, $k_B$ is the Boltzmann constant, $T$ the system temperature, and $D$ a dimensional constant. Finally, the quantity $c_J(t)=\langle J(t^\prime)J(t^\prime-t) \rangle$ is the time correlation function of the energy current $J(t)$, defined as
\begin{equation}
J(t)={1\over 2}\sum_{i=1}^N ({\dot u}_{i+1} + {\dot u}_i) \, F_i.
\end{equation}
In addition, one also can define in the same fashion the momentum current as
\begin{equation}
J_P(t)=\sum_{i=1}^N F_i.
\end{equation}
For computational reasons, it is convenient to work in the Fourier space. By taking the modulus square of the time-Fourier transform of the energy density current we define
\begin{equation}
C_\mathcal{E}(\omega)=\langle|\hat J(\omega)|^2\rangle,
\end{equation} 
where $\langle...\rangle$ denotes the average over a set of independent molecular dynamics runs. 
In the linear response regime, an estimate of 
the exponent $\gamma$ is obtained by fitting the low-frequency part of 
the spectra by an inverse power-law,  $C_\mathcal{E}(\omega) \propto \omega^{-\gamma}$
\cite{2003PhR...377....1L}. 
Note that the equivalent quantities $C_P(\omega)$ and $C_{\rho}(\omega)$ associated to the momentum 
\begin{figure*}
\begin{center}
\includegraphics[width=0.6\textwidth]{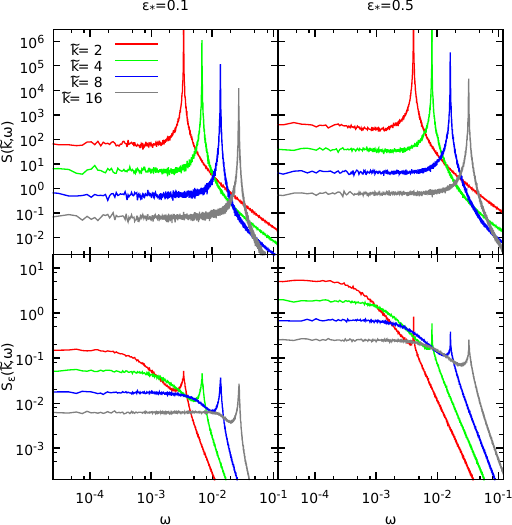}
\end{center}
\caption{(Colour online) FPU chain: Dynamical structure factor of the displacements (upper panels), and local energy (lower panels) for $\alpha = 0.1$, $N = 4096$ for different normalized wave numbers $\tilde{k} = 2$, 4, 8, and 16 and $\mathcal{E}_*=0.1$ (left), and 0.5 (right). The curves are averaged over 1000 independent molecular dynamics runs.
}
\label{structfact}
\end{figure*}
and density (displacement) currents are  constructed in a similar fashion.\\
\indent In the spirit of the NFH theory, it is also of interest to study 
dynamical scaling of the different modes of structure factors. In order to do so, we first perform the discrete space-Fourier transform of 
the particles displacement,
\begin{equation}
\hat u({k},t)=\frac{1}{N}\sum_{l=1}^Nu_l\exp(-\imath {k}l).
\end{equation}
The dynamical structure factor $S(k,\omega)$ is defined as the modulus squared of the subsequent temporal Fourier transform and reads 
\begin{equation}\label{somega}
S({k},\omega)=\langle|\hat u({k},\omega)|^2\rangle.
\end{equation}
The energy structure function $S_\mathcal{E}({k},\omega)$ is obtained in the same way after Fourier-transforming the energy density profile defined by Eq.(\ref{locen}) as
\begin{equation}\label{sene}
S_{\mathcal{E}}({k},\omega)=\langle|\hat\mathcal{E}({k},\omega)|^2\rangle.
\end{equation}
Since we are working with periodic boundary conditions, the allowed values of the wave number $k$ are always integer multiples of $2\pi/N$, therefore in the rest of the paper 
we will sometimes refer to the (integer) normalized wave number $\tilde{k} = kN/2\pi$.
\subsection{Numerical simulations and results}
\begin{figure}
\begin{center}
\includegraphics[width=0.85\columnwidth]{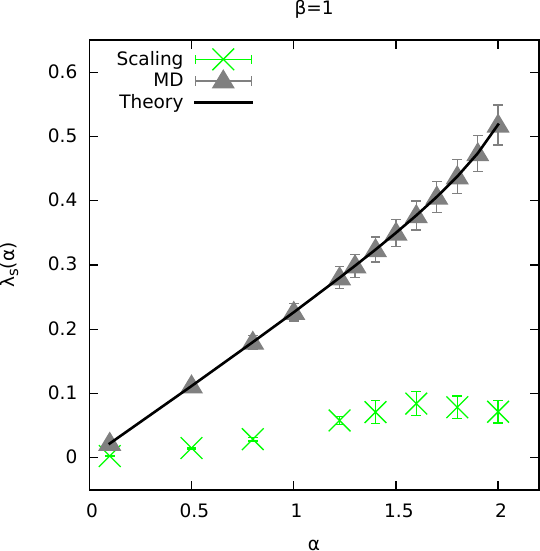}
\caption{Comparison of $\lambda_s$ obtained from dynamical scaling (crosses), the MD simulations (triangles) and the NFH theory (solid line). The values of $\lambda_s$ from the scaling have been obtained as averages over the single $k$ modes and the errors have been calculated accordingly. For this data, all simulations have been performed for $N= 1024$ and $\mathcal{E}_*=0.1$.}
\label{fig:lambda_comparison}
\end{center}
\end{figure}

\begin{figure}
\begin{center}
\includegraphics[width=\columnwidth]{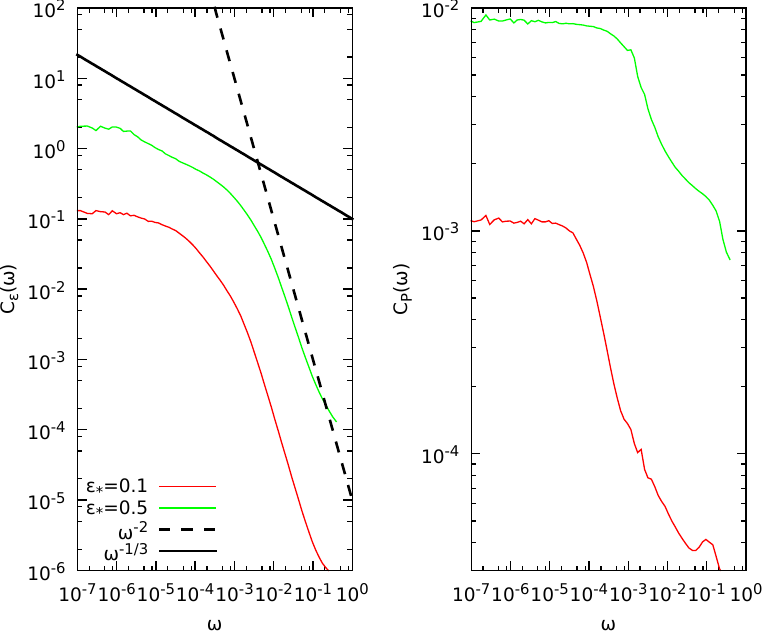}
\end{center}
\caption{(Colour online) Fourier spectra of the total energy and velocity currents $C_\mathcal{E}$ (left) and $C_P$ (right), for a FPU chain with $N=4096$, $\alpha=0.1$. Different curves refer to different values of the energy density $\mathcal{E}_*=0.1$ (lower curves) $\mathcal{E}_*=0.5$ (upper curves). Each curve is averaged over 1000 independent realizations. To guide the eye, the dashed and solid black curves with the two slopes $-1/3$ and $-2$ have been added to the plot.}
\label{corrfunctfpu}
\end{figure}

To simulate the system described above, we solve Eqs. (\ref{edm}) with a numerical code using a fourth-order symplectic integrator \cite{1995PhyS...51...29C}. The simulations presented here are {\it microcanonical} (i.e., the system is isolated and energy is, in principle, conserved) and for
fixed $\alpha=0.1$. The conservation of energy and momentum was monitored during a set of preliminary runs and we find that for $N>100$, with a timestep $\Delta t = 0.01$, the total energy of the system is conserved in the worst case up to a few parts per million. The initial conditions are implemented as follows. The $N$ particles of unitary mass are set at their equilibrium positions, and the initial velocities are extracted from a Gaussian distribution of unitary width and rescaled by a suitable factor to assign the desired value of the system specific energy per particle $\mathcal{E}_*$. Usually, a little readjustment is needed in order to set the total initial momentum $P_{\rm tot}=\sum_i \dot u_i$ equal to zero. In every case, a transient is elapsed before acquisition of statistical averages.\\ 
\begin{figure}
\begin{center}
\includegraphics[width=\columnwidth]{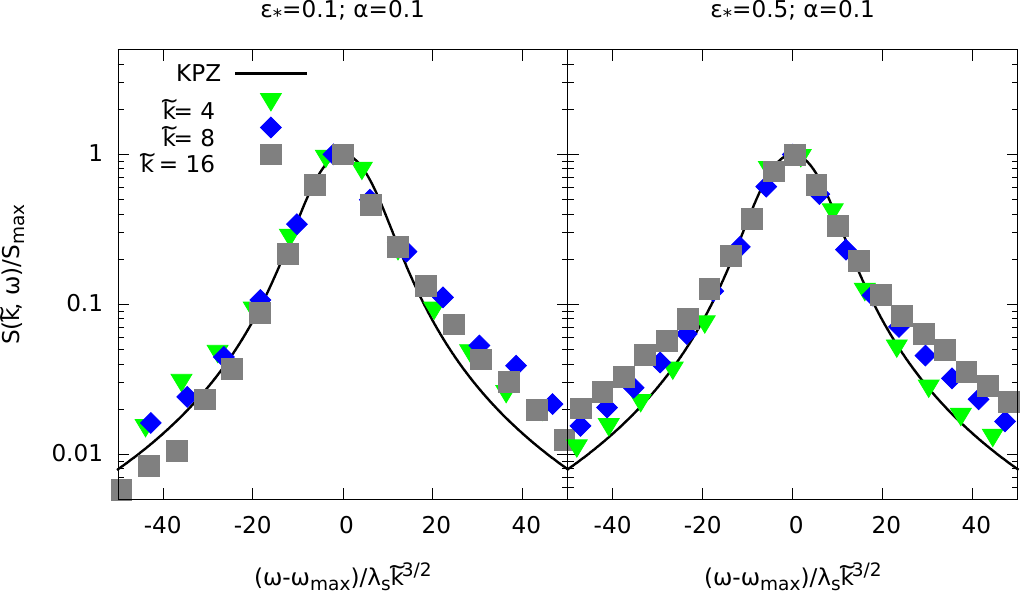}
\end{center}
\caption{(Colour online) FPU chain: Data collapse to the KPZ scaling function (solid line) of the Fourier spectra of the dynamical variable $u$ for the modes with $\tilde{k}=4,$ 8, and 16, for $\alpha=0.1$ and $\mathcal{E}_*=0.1$ (left panel), and 0.5 (right panel).}
\label{scalingfpu}
\end{figure}
\indent Figure \ref{structfact} shows  the structure factors of displacement $S(k,\omega)$ (upper panels), and energy  $S_\mathcal{E}(k,\omega)$ (lower panels) for $N=4096$, and 
different values of the normalized wave number $\tilde{k}$ (with $\tilde{k}\ll N$)
and two values of the specific energy (i.e., $\mathcal{E}_*=0.1,$ 0.5). 
For all  values of $\tilde{k}$,  $S(k,\omega)$ shows a sharp peak, whose position is proportional to the sound speed of the ballistic modes $c_s$ as $\omega_{\rm max}=c_sk$. On the other hand, $S_{\mathcal{E}}(k,\omega)$ presents for all the explored values of the wave number a prominent peak at low $\omega$ and a sharper second peak around $c_sk$. Note that the low frequency peak appears broadened due to the logarithmic scale used in the plots.\\
\indent According to the NFH theory \cite{2014JSP...154.1191S}, correlations in the 
large-time and space scale should obey dynamical scaling of the KPZ equation.
The structure factor $S(k,\omega)$ for small enough $k$ and $\omega\approx\pm \omega_{\rm max}$
should scale with the frequency $\omega$ as
\begin{equation}\label{scaling32}
S({k},\omega)\sim h_{\rm KPZ}\left(\frac{\omega\pm\omega_{\rm max}}{\lambda_s k^{3/2}}\right).
\end{equation}
Moreover, by assuming that the low-frequency part is dominated by heat modes,
one should expect that the scaling of the energy structure factor 
$S_\mathcal{E}(k,\omega)$ for $\omega\to0$  behaves as
\begin{equation}\label{scaling53}
S_\mathcal{E}({k},\omega)\sim h_{\rm LW}\left(\frac{\omega}{\lambda_h k^{5/3}}\right),
\end{equation}
Remarkably, the scaling functions are universal and known exactly: $h_{\rm KPZ}$ and $h_{\rm LW}$ are the Fourier transforms of, respectively, the KPZ scaling function and the L\`evy characteristic function of index $5/3$ (see Refs.  \citep{2014JSP...154.1191S} for details). The function $h_{\rm KPZ}$ is 
not known in closed form but has to be evaluated numerically \cite{private} while
$h_{\rm LW}$ is, by definition, a Lorentzian. The nonuniversal coefficients 
$\lambda_s$ and $\lambda_h$ are model-dependent and can be evaluated in terms 
of static correlators \cite{2014JSP...154.1191S}. A first positive test of the 
NFH predictions for the FPU model has been reported in Ref.\cite{Mendl2013}.
\begin{figure}
\begin{center}
\includegraphics[width=0.75\columnwidth]{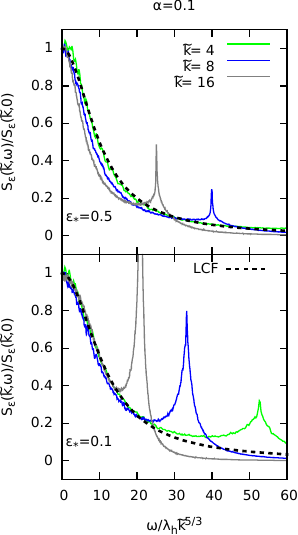}
\end{center}
\caption{(Colour online) FPU chain: Dynamical scaling for dynamical structure factors of local energy for $\alpha = 0.1$,
$N = 4096$, and $\tilde{k} = 2$, 4, 8, for values of the energy density $\mathcal{E}_*=0.1$ (bottom) and $0.5$ (top). The thick dashed line marks the Lorentzian characteristic function (LCF), as predicted by the theory. 
}
\label{termfpu}
\end{figure}
Dynamical scaling is illustrated in Fig. \ref{scalingfpu} and Fig. \ref{termfpu}
that show the data collapse to the theoretical curve for $\tilde{k}=4,$ 8 and 16.
The predicted scaling functions $h_{\rm KPZ}$ and $h_{\rm LW}$ agree very well with the observed lineshapes.
In order to test the scaling, data have been rescaled automatically, by extracting the frequency $\omega_{\rm max}$ corresponding to the maximum of the curve, and then shifting it to the origin, while the $x$ axis has been renormalized according to Eqs. (\ref{scaling32}) and (\ref{scaling53}).\\
\indent In addition, we have performed further tests of the NFH theory with focus on the nonuniversal prefactor $\lambda_s$. According to Ref.~\cite{2014JSP...154.1191S},  
$\lambda_s$ can be explicitly written in terms of  equilibrium averages of connected correlation functions of suitable thermodynamic quantities. 
In Figure \ref{fig:lambda_comparison} we report the dependence of  $\lambda_s$ on the nonlinearity parameter $\alpha$
[see Eq.(\ref{pot})]  for fixed values of the  system size ($N=1024$) and of the  specific energy ($\mathcal{E}_*=0.1$).
We compare the theoretical predictions with the values obtained for equilibrium averages, estimated by MD simulations, and the ones extracted by a fit of the structure factor $S(k,\omega)$ with the scaling function in (\ref{scaling32}).  The numerical MD results 
agree with  theoretical expectations, whereas the fitted values  deviate significantly. In particular, they underestimate systematically the theoretical values, despite the fact that 
the scaling is quite well satisfied, as shown in Figs. \ref{scalingfpu} and \ref{termfpu}. The actual peak widths of the dynamical structure factors are thus much smaller than theoretically predicted. Similar deviations in the nonuniversal coefficients were reported in Ref.~\cite{Mendl2014}.\\
\indent An unexpected result is presented in Figure 
\ref{corrfunctfpu} where we plot $C_\mathcal{E}$ and $C_{P}$ for the same 
two values of the specific energy (i.e. $\mathcal{E}_*=0.1,$ 0.5) and $N=4096$. At $\mathcal{E}_*=0.5$ the $C_\mathcal{E}$ curves show a clear $\omega^{-1/3}$ in the low frequency region before saturating. This is agreement with expectation that the model should display anomalous 
transport. However, for lower energy density, the behavior is somewhat more complicated with several slope changes, but no evidence of the low frequency singularity.
An unequivocal $\omega^{-2}$ trend for high frequencies can be observed.
Also surprisingly, the $C_{P}$ do not show any low frequency singularity
(similar saturation was observed also in Ref.~\cite{2008JSP...132....1L}).
The origin of this difference is unclear and rather puzzling in view of the 
excellent KPZ scaling observed for the structure factors for both energy values.\\
\indent In conclusion, the analysis of the structure factors as well as the test of scaling of the thermal and sound peaks points towards the success of the mode coupling theory in treating the hydrodynamics of low dimensional systems. In general, anomalous diffusion is thus expected. In addition, the apparent restoration of normal diffusion in some systems is to be interpreted only as a combined effect of finite size and (relatively) short times over which previous numerical works have been studied.   
\section{1D MPC plasma}\label{mpc}
\subsection{MPC method}
As we have anticipated in the Introduction, here we extend our study to the case of one-dimensional collisional plasma models. Several strategies exist in order to treat collisions and incorporate their contribution into standard mesh-based otherwise collisionless numerical codes. For instance, in the particle-particle-particle-mesh 
(P$^3$M) method, the usual technique used to compute the potential on the simulation grid is refined by additionally computing inside each cell the direct contribution to the force due to near particles \cite{1980CoPhC..19..215E,1981csup.book.....H}, as it is also done in Barnes-Hut tree-code \cite{1986Natur.324..446B}. Moreover, hybrid approaches are also available, where either the usual PIC scheme \cite{2000PhPl....7.3252C}, P$^3$M \cite{2014PhPl...21d3504D}, or a smooth potential method \cite{2015MNRAS.446.3150V} for the force calculation is combined with Monte Carlo sweeps in velocity space thereby restoring the collision term in the right-hand-side of the Boltzmann equation (\ref{vlasov}). In general, such methods turn out to be computationally costly since they involve iterative evaluations of the systems phase-space distribution function $f$ at each collision step.\\
\begin{figure*}
\begin{center}
\includegraphics[width=0.6\textwidth]{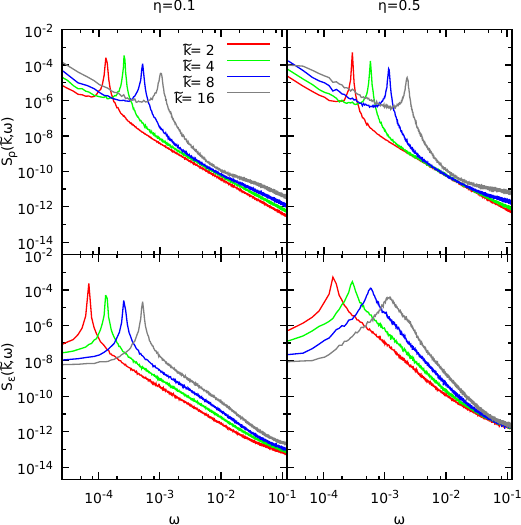}
\end{center}
\caption{(Colour online) MPC model: Dynamical structure factors  of density (upper panels) and local energy (lower panels) for $N_p = 12000$, $N_c = 1200$ for the different normalized wave numbers $\tilde{k} = 2$, 4, 8 and 16 and $\eta=0.1$ (left), and 0.5 (right). The curves are averaged over 200 independent realizations.
}
\label{structfactmpc}
\end{figure*}
\indent In this work, aiming at studying the energy transport due to inter-particle collisions in plasmas, we made use of a multiparticle collision numerical technique (MPC). The MPC method, originally introduced by Malevanets and Kapral \cite{1999JChPh.110.8605M,2004LNP...640..116M} in the context of mesoscopic dynamics of complex fluids (e.g. polymers in solution, colloidal fluids),  is based on a stochastic and {\it local} protocol that redistributes particle velocities, while preserving the global conserved quantities such as total energy, momentum and angular momentum. The algorithm (see Refs. \cite{2009acsa.book....1G,kapral08} for a detailed review) is ideally articulated in three steps:
\begin{enumerate}
 \item The system of $N_p$ particles is partitioned in $N_c$ cells where the local center of mass (c.o.m.) coordinates and velocity are computed. 
 \item Inside each simulation cell the particle velocities are rotated around a random axis passing through the center of mass, and  the rotation angles are assigned in a way that the invariant quantities are locally preserved.
 \item All particles are propagated freely, or under the effect of an external force if present.
\end{enumerate}
In this work we consider one dimensional systems with periodic boundary conditions. Therefore, inside each cell on which the system is coarse grained, the conserved quantities are the linear momentum $P_i$ and the kinetic energy $K_i$ \footnote{Here the velocity exchange is an instantaneous process that is not mediated by an effective potential, therefore we impose the conservation of the kinetic energy solely.}. During the collision step the stochastic velocity shifts  $w_j$ are extracted for each particle from a distribution depending on the cell temperature (see also Ref. \cite{tesidelfini}), and the conservation of $P_i$ and $K_i$ reads
\begin{eqnarray}\label{sist}
P_i&=&\sum_{j=1}^{N_i} m_jv_{j,{\rm old}}=\sum_{j=1}^{N_i} m_jv_{j,{\rm new}}=\nonumber\\
&=&\sum_{j=1}^{N_i} m_j(a_iw_j+b_i);\nonumber\\
K_i&=&\sum_{i=1}^{N_j} m_j\frac{v_{j,{\rm old}}^2}{2}=\sum_{j=1}^{N_i} m_j\frac{v_{j,{\rm new}}^2}{2}=\nonumber\\
&=&\sum_{j=1}^{N_i} m_j\frac{(a_iw_j+b_i)^2}{2},
\end{eqnarray}
where $N_i$ is the number of particles in cell $i$, $m_j$ and $v_j$ are the $j$-th particles mass and velocity, and $a_i$ and $b_i$ are the unknown cell-dependent coefficients. Equations (\ref{sist}) constitute a linear system that can be solved for $a_i$ and $b_i$.
\begin{figure}
\begin{center}
\includegraphics[width=\columnwidth]{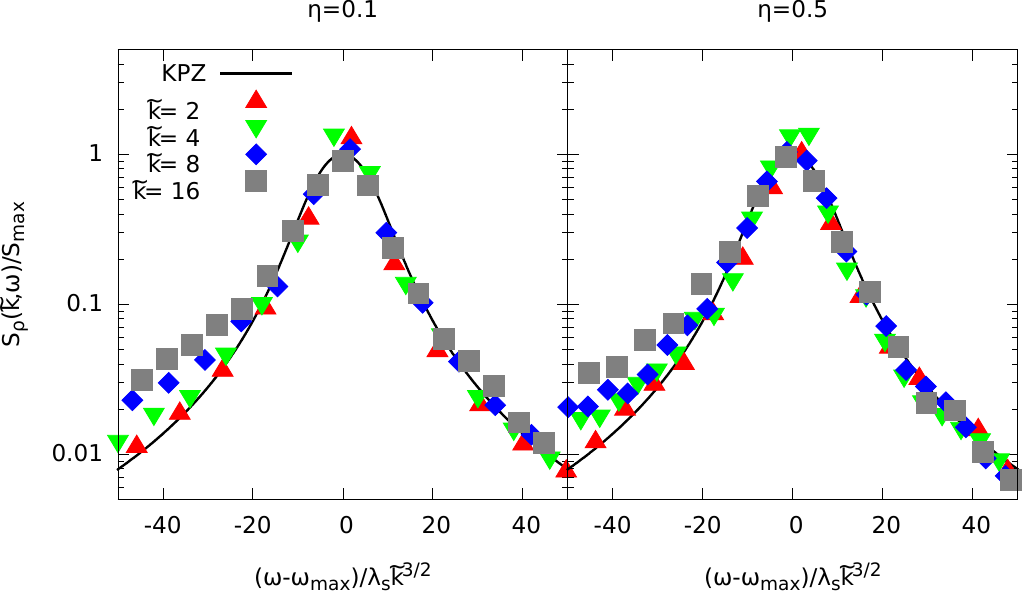}
\end{center}
\caption{(Colour online) MPC model: Data collapse to the KPZ scaling function (solid line) of the Fourier spectra of the density profile modes with $\tilde{k}=2,$ 4, 8, and 16, for $\eta=0.1$ (left panel), and 0.5 (right panel). The data is from the same calculations presented in Fig. \ref{corrfunct} (red and green curves).}
\label{scaling}
\end{figure}
First of all we introduce the auxiliary quantities
\begin{eqnarray}
P_i^*=\sum_{j=1}^{N_i} m_jw_{j}; \quad K_i^*=\sum_{j=1}^{N_i} m_j\frac{w_{j}^2}{2},
\end{eqnarray}
and then rescale them, together with $P_i$ and $E_i$, by the total mass in cell $i$,  $M_i=\sum_{j=1}^{N_i} m_j$:  
\begin{eqnarray}\label{sist4}
\tilde{P_i^*}=P_i^*/M_i;\quad\tilde{P_i}&=&P_i/M_i;\nonumber\\
\tilde{K_i^*}=K_i^*/M_i;\quad\tilde{K_i}&=&K_i/M_i.
\end{eqnarray}
Once setting
\begin{equation}
\sigma_i=\sqrt{2\tilde K_i-\tilde P_i^2};\quad\sigma_i^*=\sqrt{2\tilde{K_i^*}-\tilde{P_i^{*2}}},
\end{equation}
after easy algebra, the coefficients $a_i$ and $b_i$  are obtained as
\begin{equation}
a_i=\sigma_i/\sigma_i^*;\quad b_i=\tilde P_i-\tilde{P_i^*}a_i,
\end{equation}
and the new velocities then read
\begin{equation}
 v_{j,{\rm new}}=a_iw_j+b_i.
\end{equation}
In the propagation step the positions $r_j$ are updated and at the next step a new partitioning of the system is operated, and the procedure repeats.\\
\indent So far we have been discussing the MPC method in the general case. In order to adapt such general scheme to the modelization of plasmas of charged particles interacting with Coulomb forces, the velocity sweep protocol has to be conditioned to a local probability capturing the essence of the Coulombian scattering at low impact parameters (i.e. of the order of the cell size). A test particle of mass $m$ and charge $Q$, moving at velocity $\mathbf{v}$ in a homogeneous background of particles of the same kind, with number density $n$ and Maxwellian velocity distribution, experiences on average in the time interval $\delta t$ a number of collisions $N_{\rm coll}=\delta t\omega_{\rm coll}$, where the collision frequency scales as
\begin{equation}\label{omegacoll}
\omega_{\rm coll}\propto\frac{8\pi Q^4 n\ln\Lambda}{m^2|\mathbf{v}|^3}.
\end{equation} 
In the expression above the quantity $\ln\Lambda$ is the so-called Coulomb logarithm of the maximum and minimum impact parameters $b_{\rm max}$ and $b_{\rm min}$, whose definitions are somewhat arbitrary. Usually, in a neutral plasma $b_{\rm max}$ is the Debye screening length and $b_{\rm min}$ the minimum inter-particle distance \cite{1965pfig.book.....S}. Since one has to define a cell dependent interaction probability proportional to the local (average) collision frequency, it is tempting to use a cell-averaged collision frequency $\langle\omega_{\rm coll}\rangle$, which requires a proper rescaling of the time units so that the product $\Delta t\langle\omega_{\rm coll}\rangle$ can be used as an effective collision probability. However, given that the local velocity distribution may in principle not be Maxwellian, and since the present model is intended to capture the essence of the problem, in the simulations presented here, we condition instead the interaction step to the cell dependent Coulomb-like {\it interaction probability} \cite{2013PhRvE..87b3102B}
\begin{equation}\label{prob}
 \mathcal{P}_i=\frac{1}{1+(\tilde K_i/\mathcal{E}_{\rm int})^2},
\end{equation}
where $\mathcal{E}_{\rm int}$ is a typical interaction energy per unit mass, proportional to $n^{1/3}\langle Q^2\rangle/\langle m\rangle$, where the mass and the (squared) charge are averaged over the different component of the system. With such a choice, in a system with a given number density $n$, $\mathcal{E}_{\rm int}$ becomes a scale quantity of the simulations that can be normalized to unity. Note that, a system with two or more species of particles
might have, in principle, more complex transport properties
since the mean collision frequency strongly depends on the mass (or charge) spectrum, see e.g. Ref. \cite{2010AIPC.1242..117C}. Test simulations in which
a monocomponent plasma with initially nonthermal velocity
distribution relaxes to thermal equilibrium have been performed with standard molecular dynamics and MPC, both approaches yielding similar results.\\
\indent In order to include in the model the effects of the long-range part of the Coulomb interaction, we could rely on conventional particle-mesh schemes solving the Poisson and Amp\`ere equations on the grid to compute the electrostatic potential and magnetic field. In this work however, we do not account for these self-consistent fields.
\subsection{Numerical simulations and results}
\begin{figure*}
\begin{center}
\includegraphics[width=0.8\textwidth]{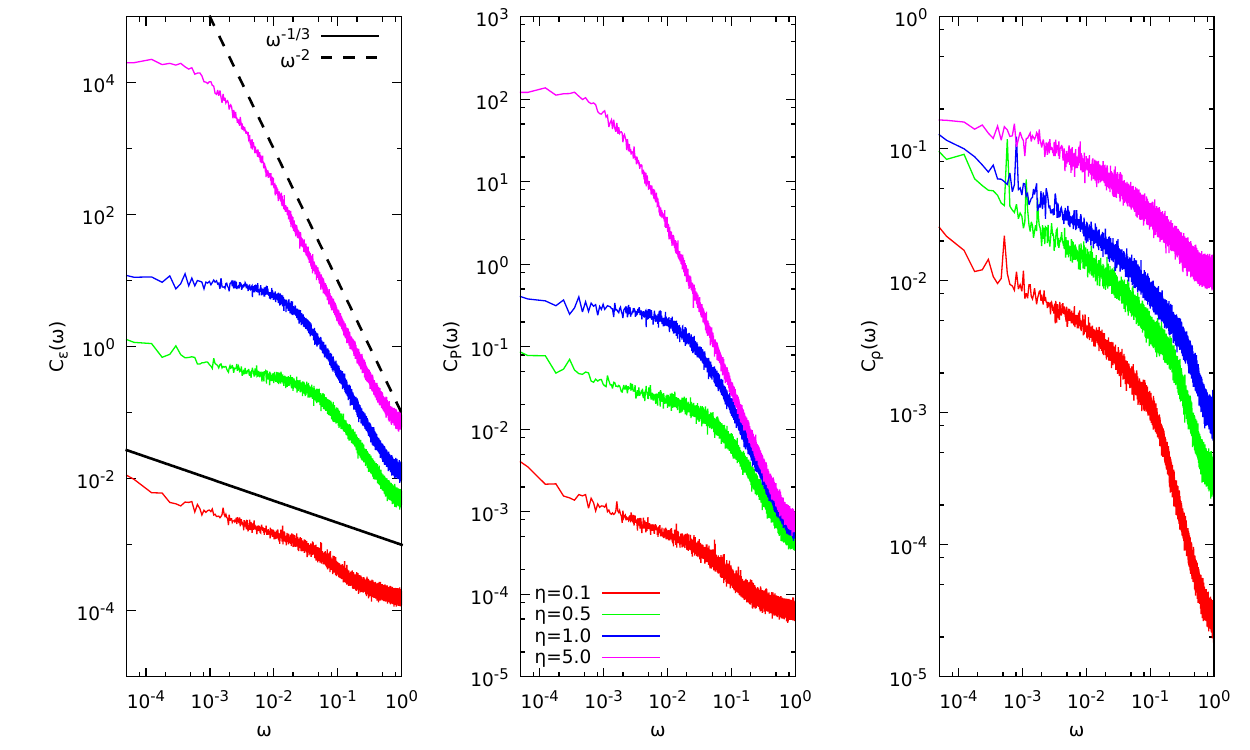}
\end{center}
\caption{(Colour online) MPC model: From left to right, Fourier spectra $C_\mathcal{E}$, $C_{P}$ and $C_{\rho}$ of the energy, momentum and density currents, for $\eta=0.1,$ 0.5, 1 and 5, and $N_c=1200$. The curves are averaged over 200 independent realizations. The cross-over from the $\omega^{-1/3}$ to the $\omega^{-2}$ behavior of $C_\mathcal{E}$ at around $\eta=0.5$ is evident. To guide the eye, the dashed and solid black curves with the two slopes $-1/3$ and $-2$ have been added to the plot.}
\label{corrfunct}
\end{figure*}
In the line of previous works \cite{1975PhRvA..12.1106V,PhysRevA.18.2345,1985PhRvA..32.2981M,1998PhRvL..81.1622D,2003PhPl...10.1220S,2015CoPP...55..421K}, we investigate the transport properties of a one component plasma via its dynamical structure factors. Since the scope of this paper is to study and compare transport in low-dimensional models, we limit ourselves to consider only one dimensional plasmas in a static neutralizing background. In analogy with the analysis of the FPU lattice, we computed for the plasma-like model the structure factors of kinetic energy, momentum and density $C_\mathcal{E}$, $C_{P}$ and $C_{\rho}$ defined as the time-Fourier transforms of the currents $J_\xi$, associated to the quantity $\xi$ on the simulation grid, that read
\begin{equation}
J_\xi(t)=\sum_{i=1}^{N_c} \left[ \xi_i(t)-\xi_{i-1}(t-\Delta t)\right].
\end{equation}
The equilibrium initial conditions are implemented as follows: $N_p$ identical particles are homogeneously placed on the simulation grid and their charges $Q$ and masses $m$ are chosen so that $\mathcal{E}_{\rm int}=1$. The initial particles velocities are extracted from a Maxwellian distribution, adjusted in order to have vanishing total momentum, and renormalized to obtain the wanted value of the specific kinetic energy per unit mass $\mathcal{E}_{\rm b}$. With this choice we have to tune a single control parameter, the ratio $\eta=\mathcal{E}_{\rm b}/\mathcal{E}_{\rm int}$, that defines the strength of the coupling between particles in the plasma implemented via Equation. (\ref{prob}).\\
\indent The particles' equations of motion are integrated in the propagation step by a second order symplectic scheme with fixed $\Delta t$. Test simulations activating the self-consistent electrostatic field yield $\Delta t=1/100\omega_P$ as the optimal value for the timestep, ensuring energy conservation up to roughly one part in $10^5$, $\omega_P=\sqrt{nQ^2/m}$ being the plasma frequency of the system in computational units (for the typical simulation parameters used here $\omega_P\approx 3$).\\
\indent Figure \ref{structfactmpc} shows the structure factors of density and energy for two strongly collisional cases with $\eta=0.1$ and 0.5.\\
\indent Contrary to the analogous plots for the FPU systems, at fixed wave number the peak of the density structure factor (corresponding to that of displacement in the case of FPU) and the ballistic peak in the energy structure factor are not placed at the same frequency $\omega$ (proportional to the sound speed). This is mainly due to the fact that in fluid models, contrary to solid, heat transfer is also due to actual mass transport, since elements of fluid can overtake each other, and the same energy $\delta\mathcal{E}$ can be transported between two parts of the system by either a few energetic particles, or a lot of fewer energetic particles. In solid models of oscillators with only nearest-neighbor couplings, on the other hand, the fluctuations of the displacement $u$ are proportional to those of particles energy due to the potential energy $V$ being a function of $u$.\\
\indent Additionally, we found that as for the FPU system, the dynamical structure factors of the density $S_{\rho}$ nicely fit with the KPZ-scaling function for all explored values of $\eta$. In Figure \ref{scaling}, we show the rescaled curves for $\eta=0.1$ and 0.5 (i.e. highly collisional systems) and $\tilde{k}=2,$ 4, 8, 16. Even for a fluid-like model such as MPC the predictions of NFH hold true for combinations of parameters associated to both anomalous or diffusive transport on the time scale of the simulations. This reinforces the idea that the apparent restoration of the normal conductivity is a non-asymptotic effect.\\
\indent In Figure \ref{corrfunct} we present the Fourier spectra $C_\mathcal{E}$, $C_{P}$ and $C_\rho$ of the energy, momentum \footnote{Note that, since all particles have the same mass $m$, we have computed the velocity current instead of the momentum current.} and density currents, respectively, for four typical values of the ratio $\eta=0.1,$ 0.5, 1 and 5, and for $N_p=12000$ particles distributed on $N_c=1200$ cells. Each simulation is extended up to $t_f=2^{19}\Delta t$. For strongly interacting systems (i.e. $\eta\leq 0.1$) one recovers the $\omega^{-1/3}$ behavior of the energy correlator $C_{\mathcal{E}}$. Increasing the particle specific kinetic energy at fixed $\mathcal{E}_{\rm int}$ (i.e. reducing the collisionality of the system), $C_\mathcal{E}$ shows a more and more prominent flat region at low frequencies departing form the $\omega^{-1/3}$ trend, and a high frequency tail with slope $\omega^{-2}$. This fact could be naively interpreted as the restoration of normal conductivity. However, one has to bear in mind that the curves are plotted over the {\it same} frequency interval. It is therefore only a finite time effect induced by the longer relaxation times of the fluctuations, due to the lower coupling in this regime. The cross-over from the $\omega^{-1/3}$ to the $\omega^{-2}$ behavior of $C_\mathcal{E}$ is evident at around $\eta=0.5$. Remarkably, the Fourier spectrum of the momentum current $C_{P}$ (central panel, same figure) has the same slope of $C_\mathcal{E}$ at fixed $\eta$. A different behavior is instead found for the density correlator $C_{\rho}$, showing instead a $\omega^{-0.45}$ slope in the central part and a $\omega^{-2}$ tail at large $\omega$.
\section{Conclusions and perspectives}
\label{discussion}
We have compared the anomalous transport properties emerging in 1D models
of a nonlinear solid, namely the FPU-chain (with leading cubic 
nonlinearity),  and a gas of particles subject to an effective Coulomb interaction.
On the basis of (NFH) \cite{2012PhRvL.108r0601V}, both models belong to the same 
universality class, namely their heat conductivities are expected to exhibit a
power-law divergence with the system size as $\kappa \sim N^{1/3}$. 
Our study confirms that for both models the
scaling properties predicted by NFH are very well recovered for both the heat and sound 
modes structure factors, in a wide range of energies.\\ 
\indent On the other hand, our analysis has unveiled discrepancies concerning the numerical results and theoretical predictions of the nonuniversal scaling coefficient $\lambda_s$. Moreover, some significant deviations and crossovers have been observed for some 
parameters in the current correlators, which decay much faster than predicted. 
This should be compared with Refs. \cite{2013PhRvE..87c2153C} 
that challenged the predictions of the NFH theory by
claiming that thermal conductivity could turn to a normal behavior in the low-energy
region of oscillators chain. Our results, along with Refs.~\cite{2014JSP...154.1191S,2015JSP...tmp...48S}, 
suggest instead that this puzzling phenomenon
should be attributed to dramatic finite--size and --time effects, rather than assuming that 
normal conduction should characterize the asymptotic transport properties of lattices
with asymmetric interaction potentials. 
This is particularly evident for the MPC gas where
structure factors again exhibit the scaling predicted by NFH over a wide range of values
of the control parameter (the interaction energy $\mathcal{E}_{\rm int}$)
whereby a clear crossover is seen in the current spectra upon reducing 
the collisionality of the particles
(see again the first panel of Fig. \ref{corrfunct}). However, the physical origin of the effect is yet unexplained. It may be traced back to  spontaneous localized density fluctuations typical of nonlinear solid models, where particles are not allowed to overtake each other or, more generally, from other slow excitations non included in the 
FHD description.
It is also to be understood how this should affect some correlation functions but not 
others. In addition, it must be pointed out that for symmetric inter-particle potentials (e.g FPU-$\beta$) a different scaling from the KPZ one is expected, as well as higher values of the exponent $\gamma$ \cite{Lepri03,2011JSMTE..03..028P,2015arXiv150505987S}.\\
\indent Although the MPC model was introduced here to test the NFH prediction in 1D, 
we point out that such a protocol can be easily 
extended to a wide range of transport problems in any dimension.
In particular, it should be mentioned that in both plasma physics and stellar dynamics there are examples of systems in which the contribution of the particle collisions to their dynamical evolution and transport properties is non negligible. For example, in the context of the frictional cooling of charged particle beams \cite{2004NIMPA.524...27G,2012PhRvS..15b4003G}, whenever an ion beam is injected into a denser electron plasma, Coulomb collisions with the background electrons have the effect of decelerating the beam particles along the initial direction of propagation (a phenomena referred to in stellar dynamics as {\it dynamical friction} \cite{1943ApJ....97..255C}). Moreover, in tokamak (generally collisionless) plasmas, whenever a hot region is connected to the colder wall, a strong temperature gradient appears. In this case, the properties 
of the plasma and therefore its collisionality vary strongly along the temperature gradient affecting the plasma's transport properties \cite{2000SPP.....7.....S}. On the side of gravitational systems, the dynamics in galaxy cores around massive central black holes is dominated by two-body encounters with low impact parameter \cite{2006RPPh...69.2513M,2015ApJ...804...52M}, while in the rest of the galaxy the dynamics is collisionless over times of the order of the age of the Universe. In all these cases, a consistent numerical treatment is needed.\\ 
\indent As a natural follow-up to this work we are going to simulate similar
gas dynamics in the presence of a self-consistent electrostatic potential
for systems with different species of particles -- a problem of primary interest
for transport phenomena in plasmas. Furthermore, we are planning to introduce 
heat and particle reservoirs in the MPC scheme to tackle problems of 
realistic conduction in gases of neutral and charged particles.
\section*{Acknowledgements}
We thank Ph. Ghendrih, V. Popkov and H. Spohn for the stimulating discussions at an early stage of this project. Two of us (HB and GC) acknowledge that this work has been carried out thanks to the support of the A*MIDEX project (Grant No. ANR-11-IDEX-0001-02) funded by the "Investissements d'Avenir" French Government program, managed by the French National Research Agency (ANR). PFDC was partially supported by the INFN project DYNSYSMATH 2015.
\bibliography{biblio.bib}
\end{document}